\documentstyle[12pt]{article}
\newcommand{\A}{{\cal A}}
\newcommand{\G}{{\cal G}}

\newcommand{\be}{\begin{equation}}
\newcommand{\ee}{\end{equation}}
\newcommand{\wt}{\widetilde}
\newcommand{\ben}{\begin{eqnarray}\displaystyle}
\newcommand{\een}{\end{eqnarray}}
\newcommand{\refb}[1]{(\ref{#1})}

\begin{document}

{}~ \hfill\vbox{\hbox{hep-th/9512203}%\hbox{MRI-PHY/28/95}
}\break

\vskip 3.5cm

\centerline{\large \bf $T$-DUALITY OF $p$-BRANES}

\vspace*{6.0ex}

\centerline{\large \rm Ashoke Sen\footnote{On leave of absence from Tata
Institute of Fundamental Research, Homi Bhabha Road, Bombay 400005, INDIA}
\footnote{E-mail: sen@mri.ernet.in, sen@theory.tifr.res.in}}

\vspace*{1.5ex}

\centerline{\large \it Mehta Research Institute of Mathematics}
 \centerline{\large \it and Mathematical Physics}

\centerline{\large \it 10 Kasturba Gandhi Marg, Allahabad 211002, INDIA}

\vspace*{4.5ex}

\centerline {\bf Abstract}

We investigate possible existence of duality symmetries which
exchange the Kaluza-Klein modes with the wrapping modes of a BPS
saturated $p$-brane on a torus. Assuming the validity of the
conjectured $U$-duality symmetries of type II and heterotic string
theories and $M$-theory, we show that for a BPS saturated $p$-brane
there is an SL(2,Z) symmetry that mixes the
Kaluza-Klein modes on a $(p+1)$ dimensional torus $T^{(p+1)}$ with
the wrapping modes of the $p$-brane on $T^{(p+1)}$. The field that
transforms as a modular parameter under this SL(2,Z) transformation
has as its real part the component of the $(p+1)$-form gauge field
on $T^{(p+1)}$, and as its imaginary part the volume of $T^{(p+1)}$,
measured in the metric that couples naturally to the $p$-brane.

\vfill \eject

Solitonic $p$-dimensional extended objects ($p$-branes)\cite{HS,DUFF}
have played an important role in the recent developments in string
theory. Many of the BPS states in string theory that
are required to exist by various dualities\cite{HT},
can be interpreted as
wrapping modes of $p$-branes on $p$-cycles of internal manifold.
Even though at present the world-volume theory of $p$-branes is ill
understood for $p\ge 2$, $p$-branes have provided useful insight
into the non-perturbative dynamics of string
theories\cite{TOWNSEND}-\cite{MUKHI}. Recent discovery\cite{POLCHINSKI}
that $p$-branes carrying Ramond-Ramond (RR) charges have exact
conformal field theoretic description as Dirichlet branes has opened
up new avenues for testing various duality conjectures in string theory,
and at the same time, has provided new insight into the dynamics of
$p$-branes\cite{WITD}-\cite{DOD}.

One of the early attempts to use $p$-branes
in the study of duality symmetries in
string theory was made in ref.\cite{SCHSE}. There it was observed
that the $S$-duality symmetry of the heterotic string theory
compactified on a six dimensional torus can be interpreted as the
$T$-duality of the theory of five-branes\cite{DU5,STR5}, in the sense
that it exchanges the Kaluza-Klein modes with the winding modes of
the five-brane on the 6-torus. Subsequently, attempt was made,
following earlier work of Duff and Lu\cite{DUFFLU}, to show that
this is a symmetry of the 5-brane world-volume theory, but this
program ran into difficulties\cite{SEZGIN,UNPUB}.\footnote{Later
$S$-duality of the heterotic string theory was realised as the
$T$-duality of the dual type IIA string theory\cite{DUS,WIS}.}
By now, however,
there has been mounting evidence for $S$-duality of heteroric
string theory compactified on a 6-torus. Thus we can now turn the
argument around and say that the complete theory has a symmetry which
acts as the $T$-duality transformation of the 5-brane, even though
the 5-brane world-volume theory itself may not be invariant under
this transformation. This result might, in fact, point to a
reformulation of the five-brane world-volume theory where this
$T$-duality symmetry will be manifest.

In this paper we shall use similar reasoning to study the possible
existence of $T$-duality symmetry
that exchanges the Kaluza-Klein modes with
the wrapping modes of a $p$-brane for general BPS saturated
$p$-branes. Our starting point will be the various conjectured
$U$-duality symmetries in various string theories. We shall assume
these $U$-duality conjectures to be true, and then try to
re-interprete some of the $U$-duality transformations as $T$-duality
transformations on various $p$-branes.

For fundamental strings, $T$-duality transformation that inverts
the volume of the internal torus, exchanges the
Kaluza-Klein modes with the winding modes of the string. This makes
sense, since for compactification on a $k$-dimensional torus, both
the winding number vector and the momentum vector are $k$-dimensional,
with integer entries. However, for $p$-branes the situation is
somewhat more complicated. The wrapping modes of a $p$-brane on a
$k$-torus ($k\ge p$) is described by a ${k \choose p}$ dimensional
vector, whereas the Kaluza-Klein modes are still described by a
$k$-dimensional charge vector. Thus it does not make sense to talk
about a symmetry that exchanges the two vectors. However, the
dimensionality of the two vectors coincide for $k=p+1$, and hence in
this case such an exchange symmetry does make sense. This is the case
that we shall focus on in this paper. Our result can be stated as
follows:

{\it
For a BPS saturated $p$-brane, invariant
under half of the global supersymmetry generators
in a $D$-dimensional theory
(obtained by compactification of string/$M$- theory) with at least
16 global supersymmetry generators,\footnote{For theories
with less number of supersymmetries even the $T$-duality transformations
involving fundamental strings get modified by quantum
corrections\cite{KAPL,NARAIN}.}
there appears
an SL(2,Z) symmetry when we compactify the theory on a $(p+1)$
dimensional torus. Let $\A_{M_1\cdots M_{p+1}}$ denote the
$(p+1)$-form field strength in $D$-dimensions $(0\le M_i
\le D-1$)and $\G_{MN}$ denote the metric in $D$-dimensions, which
couple naturally to the $p$-brane. If $(D-p-1), \ldots (D-1)$
denote the compact directions, and $\G_{\{(D-p-1)\ldots (D-1)\}}$
denotes the determinant of the $(p+1)\times (p+1)$ matrix $\G_{mn}$
($(D-p-1)\le m,n \le (D-1)$), then
\be \label{e1}
\tau\equiv \A_{(D-p-1)\ldots (D-1)} + i \sqrt{\G_{\{(D-p-1)\ldots
(D-1)\}}}\, ,
\ee
transforms under the SL(2,Z) transformation as
\be \label{e2}
\tau\to {p\tau + q \over r\tau + s}\, ,
\ee
where
\be \label{e3}
ps-qr=1, \qquad \qquad p,q,r,s \in Z\, .
\ee
Let $p_m$ denote the momenta in the internal directions, which couple
to $\G_{m\mu}$, normalized so that $p_m$ are integers. Also, let
$w_m$ denote the winding number of the $p$-brane on $p$-cycles of
the internal torus $T^{(p+1)}$ which couple to
$\epsilon^{mm_1\ldots m_p} \A_{m_1\ldots m_p\mu}$, normalized so
that $w_m$ are integers. Then the SL(2,Z) transformation acts on
these charge vectors as
\be \label{e4}
\pmatrix{p_m\cr w_m} \to \pmatrix{p & q \cr r & s} \pmatrix{p_m
\cr w_m} \, .
\ee
The $(D-p-1)$ dimensional canonical Einstein metric remains invariant
under this duality transformation.
}

We shall now verify this result for diverse $p$-branes appearing in
type II and heterotic string theories, as well as in the $M$-theory.
Note, however, that not all of these results are independent, since
many of the $p$-branes that we shall be analyzing can be reinterpreted
as wrapping modes of higher branes on internal tori.

\noindent{\bf 1. Elementary strings in type II / heterotic string
theory on $T^2$}:

In this case the SL(2,Z) transformation described above coincides
with the usual $T$-duality transformation in string theory.

\noindent{\bf 2. Neveu-Schwarz (NS) 5-brane in Heterotic / Type IIA
theory compactified on $T^6$}:

This case has been discussed in ref.\cite{SCHSE} and the resulting
SL(2,Z) coincides with the $S$-duality transformation in the four
dimensional string theory. We shall work out this case here in some
detail.

Let us denote by $G^{(NS5)}$ the metric that couples to the
Neveu-Schwarz five-brane, and by $G^{(ST)}$ the metric that
couples to the fundamental string. Also. let $\Phi$ be the ten
dimensional dilaton. Then\cite{DUFFLU1}
\be
G^{(NS5)}_{MN} = e^{-{2\Phi\over 3}} G^{(ST)}_{MN}\, .
\ee
(We are choosing a normalization convention where the string theory
effective action involving NS sector fields
is multiplied by a factor of $e^{-2\Phi}$ when written
in terms of the string metric.)
One can check the correctness of the above equation by noting that
for $G^{(ST)}_{MN}=\eta_{MN}$, the 5-brane world-volume is proportional
to $e^{-2\Phi}\sim 1/g_{st}^2$; as expected for a soliton carrying
magnetic charge under an NS sector gauge field.
In this case the 6-form field $\A$ can be identified to the field
$\wt B$ that is dual to the two form field $B_{MN}$ (in the sense
that $d \wt B \sim ~^* (d B)$, where $~^*$ denotes Hodge dual). Thus
\ben \label{e6}
\tau & = & \wt B_{456789} + i \sqrt{G^{(NS5)}_{\{456789\}}}\nonumber \\
& = & \wt B_{456789} + i e^{-2\Phi} \sqrt{G^{(ST)}_{\{456789\}}}\, .
\een
Thus $Re(\tau)$ can be identified with the four dimensional axion
field, and $Im(\tau)$ can be identified to $\exp(-2\Phi^{(4)})$, where
$\Phi^{(4)}$ is the four dimensional dilaton field. This shows
that $\tau$
is precisely the $S$ field of the four dimensional theory. In other words,
the $T$-duality of the five brane can be identified to
the $S$-duality of the four dimensional theory. As shown in
ref.\cite{SCHSE}, the transformation laws \refb{e4}
of $p_m$ and $w_m$ also
coincide with the corresponding transformation laws of the charges
under $S$-duality transformation.

\noindent{\bf 3. Supermembrane in 11 dimensional supergravity
compactified on $T^3$:}

In this case the field $\A$ corresponds to the three form gauge
potential $C_{MNP}$ of the 11 dimensional supergravity, and the
metric $\G$ corresponds to the metric $G^{(SG)}$ of the supergravity
theory. Thus
\be \label{e7}
\tau = C_{89(10)} + i \sqrt{G^{(SG)}_{\{89(10)\}}}\, .
\ee
We can regard this as a type IIA theory in (9+1) dimensions, spanned
by the coordinates $x^0, \ldots x^9$, compactified on a two
dimensional torus spanned by the coordinates $x^8$ and $x^9$.
The relationship between the fields in the supergravity theory
and the type II theory are given by\cite{WIS}
\be \label{e8}
C^{(SG)}_{89(10)} = B_{89}\, ,
\ee
and
\ben \label{e9}
G^{(SG)}_{(10)(10)} & = & e^{4\Phi\over 3} \, , \nonumber \\
G^{(SG)}_{MN} & = & e^{-{2\Phi\over 3}} G^{(ST)}_{MN}
\qquad \hbox{for} \qquad 0\le M,N\le 9\, .
\een
Thus,
\be \label{e10}
\tau=B_{89} + i \sqrt{G^{(ST)}_{\{89\}}}\, .
\ee
This shows that $\tau$ is the usual modular parameter associated with
the SL(2,Z) $T$-duality transformation of type IIA
compactified on a two dimensional torus.

One can also verify that the transformation of the charges under the
$T$-duality transformation of the 11-dimensional membrane coincides
with their transformation under the usual $T$-duality transformation of
the IIA theory compactified on $T^2$. To see this note that from the
point of view of the IIA theory, $p_8$ and $p_9$ denote the
Kaluza-Klein momenta along 8 and 9 directions respectively,
whereas $p_{10}$ denotes the $A^{(1)}_\mu$ charge, where we denote
by $A^{(p)}$ the $p$-form gauge field arising in the RR sector of
the type II theory. ($p$ is even for IIB and odd for IIA.)
On the other hand, $w_9$ and $-w_8$ denote the winding number along
$8$ and $9$ directions respectively, and $w_{10}$ denotes the
$A^{(3)}_{89\mu}$ charge. Under the SL(2,Z) $T$-duality
transformation of the membrane, $\vec p$ mixes with $\vec w$. This
is precisely the way the SL(2,Z) $T$-duality transformation of
the fundamental string acts on these charges. Thus we see that
in this case the $T$-duality transformation associated with the
supermembrane in the 11 dimensional supergravity compactified on
$T^3$ can be identified with the usual $T$-duality transformation
of the type IIA theory compactified on $T^2$.

\noindent{\bf 4. 5-brane in 11-dimensional supergravity compactified
on $T^6$}:

In this case $\A$ can be identified with the 6-form gauge potential
$\wt C$ in the 11-dimensional theory that is dual to $C$, in the
sense that $d\wt C\sim ~^*(dC)$. $\G$ is still the metric $G^{(SG)}$
of the 11-dimensional supergravity. Thus
\be \label{e11}
\tau = \wt C_{56789(10)} + i \sqrt{G^{(SG)}_{\{56789(10)\}}}\, .
\ee
We can again regard this as the type IIA theory compactified on
$T^5$. We now have
\be \label{e12}
\wt C_{56789(10)}= A^{(5)}_{56789}\, .
\ee
Using eqs.\refb{e9}, \refb{e11} and \refb{e12} we get
\be \label{e13}
\tau = A^{(5)}_{56789} + i e^{-{\Phi}}
\sqrt{G^{(ST)}_{\{56789\}}} \, .
\ee
We want to show that the SL(2,Z) transformation acting on this modular
parameter can indeed be identified with one of the known duality
transformations in string theory. To do this let us make a string
theoretic $T$-duality transformation that inverts the radii of all
the circles labelled by coordinates $x^5, \ldots x^9$. This has the
following effects:
\begin{enumerate}

\item{It converts type IIA theory to type IIB theory.}

\item{It converts $A^{(5)}_{56789}$ to $A^{(0)}$.}

\item{It converts $e^{-{\Phi}}\sqrt{G^{(ST)}_{\{56789\}}}$
to $e^{-{\Phi}}$. }

\end{enumerate}

Thus this transformation converts $\tau$ to
\be \label{e13a}
A^{(0)} + i e^{-{\Phi}}\, .
\ee
The SL(2,Z) transformation acting on the above modular parameter is
clearly the $S$-duality transformation of the 10-dimensional type IIB
theory, which has been conjectured to be an exact symmetry of this
theory. The transformation of the various charges can also be shown
to work out correctly.
Thus we see that the SL(2,Z) $T$-duality symmetry of the
5-brane in 11-dimensional supergravity compactified on $T^6$ follows
as a consequence of the already conjectured dualities involving string
theory and supergravity theory.

\noindent{\bf 5. RR $p$-brane in 10-dimensional type II theory
compactified on $T^{(p+1)}$}:

Type II theory contains $p$-branes carrying charge under the RR
$(p+1)$-form field $A^{(p+1)}_{M_1\ldots M_{p+1}}$. $p$ is even for the
type IIA theory and odd for the type IIB theory.
Here $\G$ corresponds to the metric
$G^{(RRp)}$ that couples to the RR $p$-brane, and $\A$ corresponds
to the RR gauge potential $A^{(p+1)}$. The relationship between
$G^{(RRp)}$ and the string metric $G^{(ST)}$ is given by
\be \label{e14}
G^{(RRp)}_{MN} = e^{-{2\Phi\over p+1}} G^{(ST)}_{MN}\, .
\ee
The correctness of this equation can be checked by noting that for
$G^{(ST)}_{MN}=\eta_{MN}$, the world-volume of the RR $p$-brane is
proportional to $e^{-\Phi}$. This is in agreement with the fact
that solitons carrying RR charge have their world-volume action
proportional to $g_{st}^{-1}$. Thus now
\ben \label{e15}
\tau & = &  A^{(p+1)}_{(9-p)\ldots 9}+ i
\sqrt{G^{(RRp)}_{\{(9-p)\ldots 9\}}}\, , \nonumber \\
& = &  A^{(p+1)}_{(9-p)\ldots 9}+ i e^{-{\Phi}}
\sqrt{G^{(ST)}_{\{(9-p)\ldots 9\}}}\, .
\een
Again in order to identify the SL(2,Z) transformation on this
modular parameter with one of the known duality transformations, let
us make a string $T$-duality transformation that inverts the radii
of all the circles labelled by $x^{(9-p)}, \ldots x^9$. This
gives a final theory that is always IIB, and transforms $\tau$ to
\be \label{e15a}
A^{(0)}+ i e^{-{\Phi}}\, .
\ee
Thus the SL(2,Z) transformation on this modular parameter can again
be identified to the $S$-duality of the 10-dimensional type IIB theory.
The transformation on the charges can be shown to work out as before.

Note that formally, for $p=-1$, the SL(2,Z) $T$-duality transformation
can be directly identified with the $S$-duality of the 10-dimensional
type IIB theory. Thus in a formal sense, the $S$-duality of the type IIB
theory can be interpreted as the SL(2,Z) $T$-duality associated with
the BPS saturated $-1$ branes!

\noindent{\bf 6. 0-branes in 9-dimensional heterotic / type II theory
compactified on $S^1$:}

When we compactify the 10-dimensional heterotic or type II theory on
a circle, we get some BPS states which cannot be regarded as the
wrapping modes of a higher dimensional brane. These are the
Kaluza-Klein modes and couple to the gauge field components
\be \label{e17}
\A_M\equiv G^{(ST)}_{9 M}/G^{(ST)}_{99} \qquad \hbox{for}
\qquad 0\le M \le 8\, .
\ee
According to the
general result that we have stated about $T$-duality of $p$-branes,
we should get an SL(2,Z) duality symmetry when we further compactify
the theory on a circle, which will exchange the BPS states carrying
$\A_M$ charges with the
states carrying momenta along the new compact direction. From the way
it acts on the charges, it is clear that this SL(2,Z) is nothing but
the SL(2,Z) representing the global diffeomorphism group of the 2-torus.
We shall now see that the modular parameter, calculated according to
the prescription given before, works out correctly. For this we note
that the metric $\G_{MN}$ that couples naturally to the Kaluza-Klein
modes is given by,
\be \label{e16}
\G_{MN} = (G^{(ST)}_{99})^{-1} (G^{(ST)}_{MN}
-G^{(ST)}_{M9} (G^{(ST)}_{99})^{-1} G^{(ST)}_{N9}), \qquad 0\le M,N
\le 8\, .
\ee
The overall normalization of the metric is determined by the
requirement that
the world-line action of the Kaluza-Klein mode for
$G^{(ST)}_{MN}=\eta_{MN}$ ($0\le M,N\le 8$) must be
proportional to its mass $1/R_9\sim (G^{(ST)}_{99})^{-1/2}$.
Thus
\be \label{e18}
\tau = \A_8 + i\sqrt{\G_{88}}= {G^{(ST)}_{98}\over G^{(ST)}_{99}}
+ i {\sqrt{G^{(ST)}_{88} G^{(ST)}_{99} - (G^{(ST)}_{98})^2}
\over G^{(ST)}_{99}}\, .
\ee
This is precisely the modular parameter of the two torus labelled
by the coordinates $x^8$ and $x^9$. This shows that the SL(2,Z)
associated with the diffeomorphism group of the 2-torus also can
be interpreted as the $T$-duality associated with the 0-branes.

To summarize, we have shown through various examples
that for a BPS saturated $p$-brane
invariant under half of the global supersymmetry generators in a
theory with at least 16 global supersymmetry generators, we can
define a $T$-duality transformation that exchanges the wrapping
modes of a $p$-brane on a $p+1$ dimensional torus and the
Kaluza-Klein modes. This generalises the usual $T$-duality
transformation in string theory that exchanges the string winding
modes with the Kaluza-Klein modes, and points to a kind of
democracy between all $p$-branes\cite{TOWNSEND}, whether elementary
or composite.

I would like to acknowledge the hospitality of the Indian Institute
of Science, Bangalore, where part of this work was done.

\end{document}